\documentclass[twocolumn,showpacs,amsmath,amssymb,pre,aps]{revtex4}   %%  4-pages !!
\usepackage{graphicx}% Include figure files
\usepackage{dcolumn}% Align table columns on decimal point
\usepackage{bm}% bold math
\usepackage{color}% color text

\renewcommand{\vec}[1]{\mathbf{#1}}

\newif\ifgraph

\graphtrue             %
\begin{document}
\title{Escape kinetics of  self-propelled particles from a circular cavity}

\author{ Tanwi Debnath$^{1}$, Pinaki Chaudhury$^{1}$, Taritra Mukherjee$^{2}$,  Debasish Mondal$^{3}$ and Pulak K.Ghosh$^{2}$\footnote{E-mail: pulak.chem@presiuniv.ac.in (corresponding author)}}

\affiliation{$^{1}$ Department of Chemistry,
University of Calcutta, Kolkata 700009, India}

 \affiliation{$^{2}$ Department of Chemistry,
Presidency University, Kolkata 700073, India}

 \affiliation{$^{3}$ Department of Chemistry and Center for Molecular and Optical Sciences \& Technologies, Indian Institute of Technology Tirupati, Yerpedu 517619, Andhra Pradesh, India}

\date{\today}

\begin{abstract}
We numerically investigate the mean exit time of an inertial active Brownian particle from a circular cavity with single or multiple exit windows.  Our simulation results witness distinct escape mechanisms depending upon the relative amplitudes of the thermal length and self-propulsion length compared to the cavity and pore sizes. For exceedingly large self-propulsion lengths, overdamped active particles diffuse on the cavity surface, and rotational dynamics solely governs the exit process. On the other hand, the escape kinetics of a very weakly damped active particle is largely dictated by bouncing effects on the cavity walls irrespective of the amplitude of self-propulsion persistence lengths. We show that the exit rate can be maximized for an optimal self-propulsion persistence length, which depends on the damping strength, self-propulsion velocity, and cavity size. However, the optimal persistence length is insensitive to the opening windows' size, number, and arrangement.  Numerical results have been interpreted analytically based on qualitative arguments. The present analysis aims to understand the transport controlling mechanism of active matter in confined structures.   
\end{abstract}
\maketitle

\section{Introduction} \label{intro}

\begin{figure}[tp]
\centering \includegraphics[width=8cm]{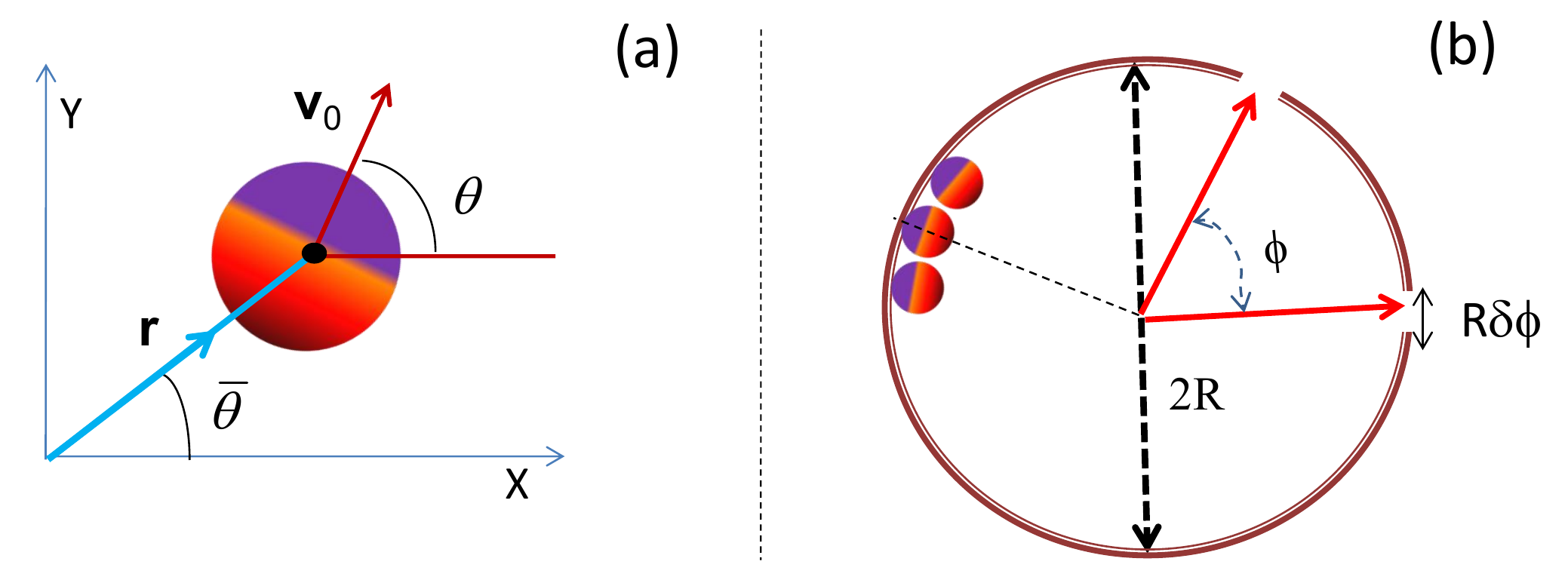}
\centering \includegraphics[width=8cm]{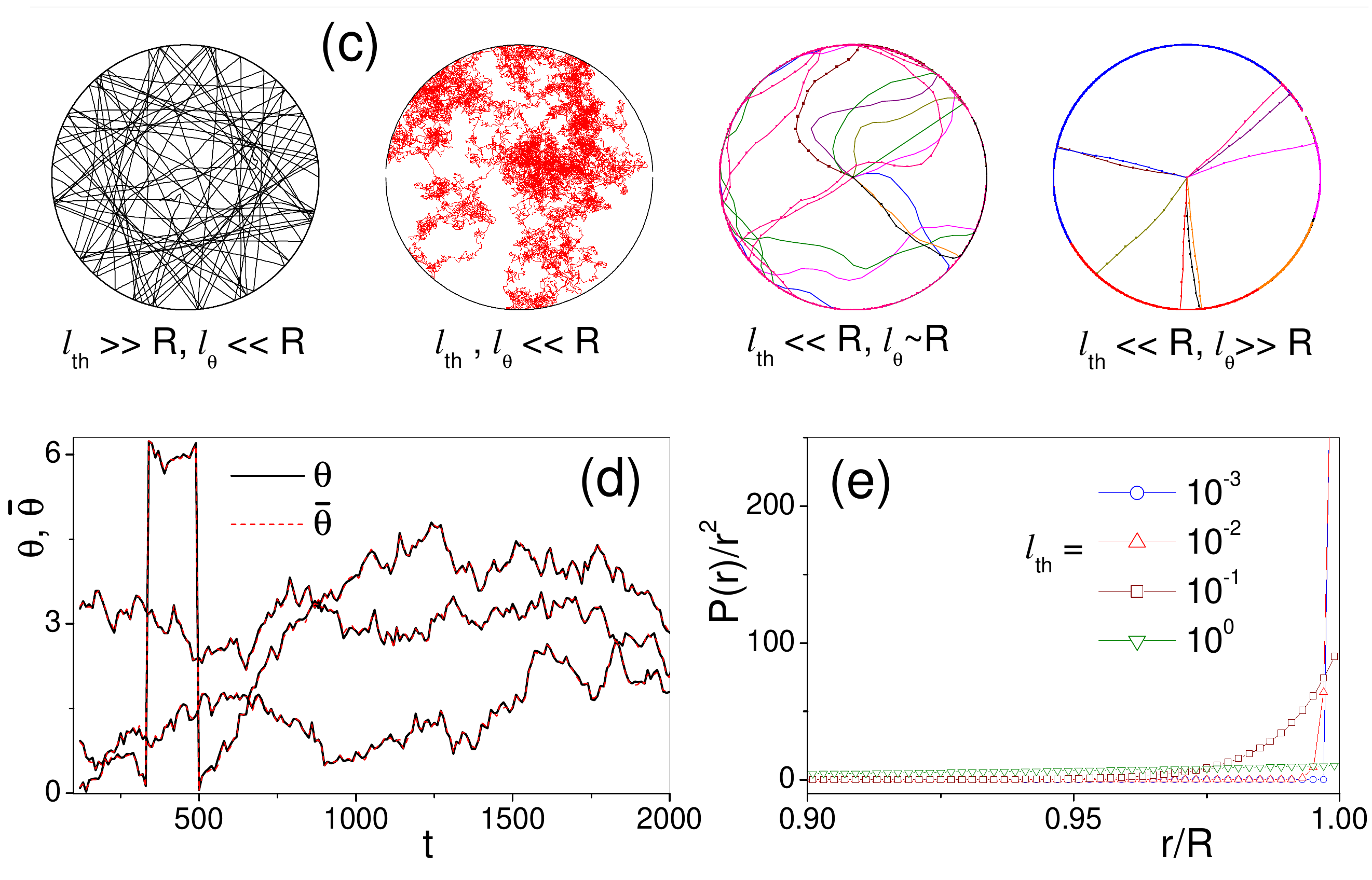}
\caption{ (Color online)  (a) Schematic represents an active Brownian particle of the Janus kind depicting orientation of the self-propulsion velocity $\vec {v_0}$  and position vector $\vec{r}\equiv (x,y)$. It is assumed that self-propulsion velocity is directed along a certain symmetry axis of the particle. $\vec {v_0}$ and $\vec {r}$ are oriented making angles $\theta $ and $\bar{\theta}$, respectively, with respect to x-axis. (b) Sketch of a circular cavity of radius $R$ with two narrow openings separated by an angle $\phi$.  Here, $R \delta \phi$ is the size of each opening. Active Brownian particles are sitting on the cavity wall when $\vec{v_0}$ is aligned to $\vec{r}$. (c) Active particle trajectories inside a circular cavity for the different regimes of $l_\theta \; {\rm and} \; l_{th}$ as shown in the legends. (d) Orientation of $\vec{v_0}$ and $\vec{r}$ with respect to x-axis as a function of time. Variations of $\theta$ and $\bar{\theta}$ are restricted within the range, 0 to $2\pi$, using their periodicity. Plots here correspond to the limit, $l_\theta >> \{R, l_{th}\}$ and $l_{th} << R $. (e) Radial probability density functions for different thermal lengths (see legends). Parameters used in (d) and (e)[unless mentioned in the legends]: $l_{\theta} = 1000, \;l_{th} = 0.001, \; R = 1, \; v_{th} = 0.1,\; v_0 = 1.$    \label{F1} } 
\end{figure}

Escape mechanisms of active species like self-propelled nano/micro-particles or living micro-organisms from entropic or energetic traps play pivotal 
role in their migration through confined structures~\cite{bacteria-diff,Marchetti,Bechinger-review,Schimansky-Geier-1} or movement on substrates~\cite{Schimansky-Geier-1}. These mechanisms help microorganisms to find strategies to explore the local environment in their natural habitat to sense stimuli. To prevent the spreading of natural micro-swimmers through porous media (like biological cellular structures  or artificial high efficiency filter media), or controlling motions of active Brownian tracer in biomedical devices or artificial nano-devices, it is desirable to have better understanding on  survival probability of particles inside the cavity as well as their escape kinetics. The escape kinetics of an active particle is considerably different from ordinary passive particles due to their time-correlated self-propulsion motion~\cite{Marchetti,Granick,Muller}.  Moreover, due to the additional 
sources of energy, active particles are typical non-equilibrium systems exhibiting peculiar transport features \cite{Golestanian-1,Golestanian-2,VolpeP,JP1,Pumera,JP2,JP2a,Misko2} and collective properties \cite{Marchetti1,Marchetti2,Redner,Buttinoni,Cates,Gompper,Stark,Misko1}.

The thermal noise-induced escape rate of passive particles from energetic traps is  well commemorated 
as Kramers' problem \cite{Kramers}. Furthermore, 
this rate theory had been extended to the external noise-driven non-equilibrium system \cite{Kramers,beyond-1,beyond-2,beyond-3,beyond-4,beyond-5,beyond-6}. Because of its success in explaining experimental findings, the Kramers' rate theory attracted wide attention over the years.

The escape dynamics of passive Brownian particles from various type of confinements and opening structures has been extensively investigated in the recent literature \cite{Zwanzig,holcman-review,jcp1,hanggi-review,Burada-review,jcp2,HangSR,ourSR1,ourSR2,Kullman,DSR1,deb1,deb3,Schimansky-Geier-2}. Usually, the exit rate from smoothly corrugated channel compartments is studied under the premises of Fick-Jacobs scheme\cite{Zwanzig,Burada-review,jcp2,HangSR}. An alternative method based on random walker schemes emerged very effective to handle cases for both smoothly as well as sharply corrugated walls~\cite{holcman-review,Kullman}. However,  both of these schemes lose their validity in case of persistent Brownian motion due to inertia or self-propulsion~\cite{jcp1,Our-inertia,JPCC-1,Schimansky-Geier-3}.

Several recent studies have been devoted to explore different aspects of the escape kinetics of self-propelled particles from metastable states~\cite{Schmid-1,jcp1,Sharma-1,Sharma-2,Tanwi-1,Ddas,JP2,Nature-exit,R5,R6,R7}. Aiming at nano-technological and biomedical applications, the escape dynamics of self-propelled particles from a compartment of entropic channels has been investigated in Refs~\cite{jcp1,JP2}. These works have focused on the impact of particle size, shape, and pore geometries on the mean exit rate, potentially important to realise transport control mechanisms in confined structures.  Geiseler et al.~\cite{Schmid-1}, studied escape kinetics of self-propelled  particles from metastable potential in the frame-work of Kramer's problem. Their study offers an analytical expression of escape rate in the small and large limits of rotational relaxation time, which can be adjusted by tuning the particle size. Based on the extensive numerical simulation, the barrier crossing dynamics has further been extended to active Brownian particles (ABPs) carrying cargo\cite{Tanwi-1}. This study shows that synchronisation between barrier crossing events and the rotational relaxation process can enhance the escape rate to a large extent. A different model of self-propulsion has been used   in Refs~\cite{Sharma-1,Sharma-2} to explore features of exit rate from a metastable state. A very recent experimental study~\cite{Nature-exit} on escape dynamics of active particles in bistable potentials shows that there exists an optimal correlation time of self-propulsion that maximises the switching rate between two wells. This feature was observed in the earlier studies for entropic as well as energetic barrier crossing dynamics of active species~\cite{jcp1,Sharma-1,Tanwi-1}.  

 Escape dynamics of overdamped active particles from circular cavities has been studied in Refs.~\cite{R1,R2,R3} under athermal conditions. These investigations, based on both run-and-tumble \cite{R1,R2} and active Brownian \cite{R2} models, noted interesting features of escape dynamics:  The mean exit rate is directly proportional to the self-propulsion persistence length in the diffusion limit. However, the rate is inversely related to the persistence length in the ballistic regime. A cross-over is observed between these two asymptotes. Indeed, all these characteristic traits of narrow escape problems are manifested even in other types of cavity structures~\cite{jcp1,JP2,R4}. Based on a minimal two-dimensional Vicsek model, the escape kinetics from a circular cavity has further been extended for interacting active particles~\cite{R3}. This study reported an exciting cross-over of survival probability from initial exponential to subexponential late-time decay.

In this paper, we revisit the escape rate of self-propelled particles from a circular cavity (see Fig.~1) with single, as well as, multiple openings. In contrast to the earlier investigations \cite{R1,R2,R3}, our study considers underdamped dynamics of self-propelled particles in a thermal environment.  Thus,  impacts of inertia and the thermal fluctuations are within the purview of the present investigation. We analyse both underdamped and overdamped limits
for three distinct regimes of self-propulsion lengths ($l_\theta$) concerning the cavity size ($R$). We show that their rotational dynamics essentially controls the escape mechanism of highly damped active particles. For prolonged rotational relaxation, particles stay on the wall most of the time, and appropriate rotational diffusion takes the particles to the exit window.   Our simulation results show that the exit rate gets maximised for an optimised persistence length of self-propulsion which strongly depends on the self-propulsion velocity, viscous relaxation time and the cavity diameter.  These results are consistent with the prior investigation in the overdamped limit \cite{R1,R2,R3}.

We show that the condition to obtain such a fastest escape 
is insensitive to the pore size, arrangement of the pores and intrinsic thermal translational motion of the particle. For a cavity with two exit windows, our theoretical predictions and numerical results show that the escape rate weakly depends on the euclidean distance between pore centres when $l_\theta >> R$. However, the opposite limit witnesses a logarithmic variation when pores are not well separated.  For swift rotational motion, active particles exhibit a similar escape mechanism as passive particles. Bouncing effects dominate the dynamics of a very weakly damped particle inside the cavity; thus, they display a distinct escape feature. Our simulation results prove that the escape kinetics weakly depends on the self-propulsion velocity and length when the thermal persistence length is larger than the cavity diameter.  Further, inertial impacts are meaningful in the dynamics as long as the thermal length is comparable to any other length scale of the escape problem. Such a situation is likely to happen when particles escape through narrow pores that are a little larger than the particle size.

The outlay of this paper is as follows. In Sec. II, we present our model that describes the dynamics of an active particle diffusing inside a cavity.  The significances of the relevant model parameters have been deliberated %discussed there 
briefly.  Numerical results are analysed in Sec. III and Sec. IV. In Sec. III, we focus on the escape kinetics of highly damped particles for distinct regimes of self-propulsion lengths. The escape dynamics of underdamped active particles has been explored in  Sec. IV. Finally, we summarise our main findings in Sec. V. %Our main findings are summarised with concluding remarks in Sec. V.  

\section{Model}\label{Model}
 We consider an inertial active Brownian particle of mass $m$ is diffusing in a 2D suspension fluid confined in a circular cavity. The active particle can exit through the openings as depicted in Fig.~1.  We model unconstrained dynamics of the active particle by the following set of equations \cite{nanoscale,Lowen-inertia1,Lowen-inertia2}:  
\begin{eqnarray}
m\ddot{x} &=& -\gamma \dot{x} + F_0 \cos{\theta}+\xi_x(t); \label{Langevin1}\\
m\ddot{y} &=& -\gamma \dot{y}+F_0 \sin{\theta} + \xi_y(t); \label{Langevin2}\\
\dot{\theta} &=& \xi_\theta(t). \label{Langevin3}
\end{eqnarray}
 This set of equations reduces to the standard active Brownian particle model~\cite{Bechinger-review,Marchetti} for the vanishingly small viscous relaxation time.    In the above equations, $\{x, y\}$ is the instantaneous position of the particle's center of mass. It is assumed that self-propulsion force with a constant modulus $F_0$ (or $v_0$) is directed along a certain particle's symmetry axis [as illustrated in Fig.~1(a)]. Further, the symmetry axis of the self-propulsion force is oriented at an angle $\theta$ with respect to the laboratory x-axis. The time evolution  of $\theta$ due to rotational diffusion of the particle is described by Eq.~(3). The random forces, $\{\xi_x(t), \;\xi_y(t), \; \xi_{\theta}(t)\}$ accounting thermal translational and rotational diffusion have been modelled by  Gaussian distributions, zero mean, and $\delta$-correlations, 
\begin{eqnarray}
\langle \xi_i(t)\xi_j (t') \rangle&=&2 \gamma k_BT \delta_{ij}\delta(t-t'); \; i,j=x,y, \label{FD}\\ 
\langle \xi_{\theta}(t)\xi_{\theta} (t') \rangle&=&2D_{\theta}\delta(t-t'), \label{rot}
\end{eqnarray} 
where, $T$, $\gamma$ and $D_\theta$ represent temperature, damping constant and rotational diffusion constant, respectively.
The equation (\ref{FD}) ensures the balance between fluctuations and dissipation; thus the particle remains in equilibrium with surrounding in the absence of self-propulsion, $\vec{F}_0 = 0$. In this limit, the particle diffuses with thermal diffusivity $D_0 = k_B T/\gamma$. The damping constant $\gamma$ determines viscous relaxation time, defined as, $\tau_\gamma = m/\gamma$. We note that the $\gamma$ is supposed to play the role of an effective viscous force taking into account all additional effects that are not explicitly accounted for in the Langevin Eq.~(\ref{Langevin1}-\ref{Langevin2}).  The thermal rotational diffusion constant for a spherical particle of radius $a$ in a viscous medium with viscosity $\eta$ is $D_\theta=k_B T/8\pi \eta a^3$. However, an active particle's rotational diffusion constant also carries non-thermal contributions that largely depend on the associated self-propulsion mechanisms. Thus, $D_\theta$ can be treated as an independent system parameter.

The components of self-propulsion act as random forces with non-Gaussian distribution and exponential correlation, $F_0^2\langle \cos\theta(t)\cos\theta(t') \rangle = F_0^2\langle \sin\theta(t)\sin\theta(t') \rangle = F_0^2\exp[-|t-t'|/\tau_\theta]$. Where, the rotational relaxation time, $\tau_\theta = 1/D_\theta$. Thus, the persistence length for unconstrained motion of the active particle is given by, $l_\theta = v_0 \tau_\theta$. Here, $v_0 = F_0/\gamma$ is the modulus of bulk self-propulsion velocity. To this end, the non-thermal active Brownian motion can be characterized by two independent parameters, either ($\tau_\theta, v_0$) or ($l_\theta,  v_0$). The effective diffusivity of an unconstrained active particles is the sum of thermal ($D_0$) and non-thermal ($D_s$) contributions. For the above model of the active particle, $D_s = v_0 l_{\theta}/2$.  

To obtain the position of the particle as a function of time, the set of Langevin equations~(\ref{Langevin1}-\ref{Langevin3}) have been numerically solved using a standard Milstein algorithm~\cite{Kloeden}. The numerical integration has been carried out using a concise time step, $10^{-4}-10^{-5}$, to ensure numerical stability. The wall-particle interactions have been modelled by the elastic reflection of velocity $\dot{r}$ at the boundary. We estimate mean exit time $\tau_{ex}$ out of the circular cavities with various initial conditions.  To be specific, $\tau_{ex}$ is defined as the average time the particle takes to exit the cavity through a narrow exit window starting with a specific initial position, velocity and direction of self-propulsion force. The mean exit times reported in this paper are obtained as averages over $10^5$ to $10^6$ trajectories.  For the simulation parameters values used here,  length and time scales are micrometres and seconds, respectively. We consider the mass of the particle to be the unit of mass.

\begin{figure}[tp]
\centering \includegraphics[width=7cm]{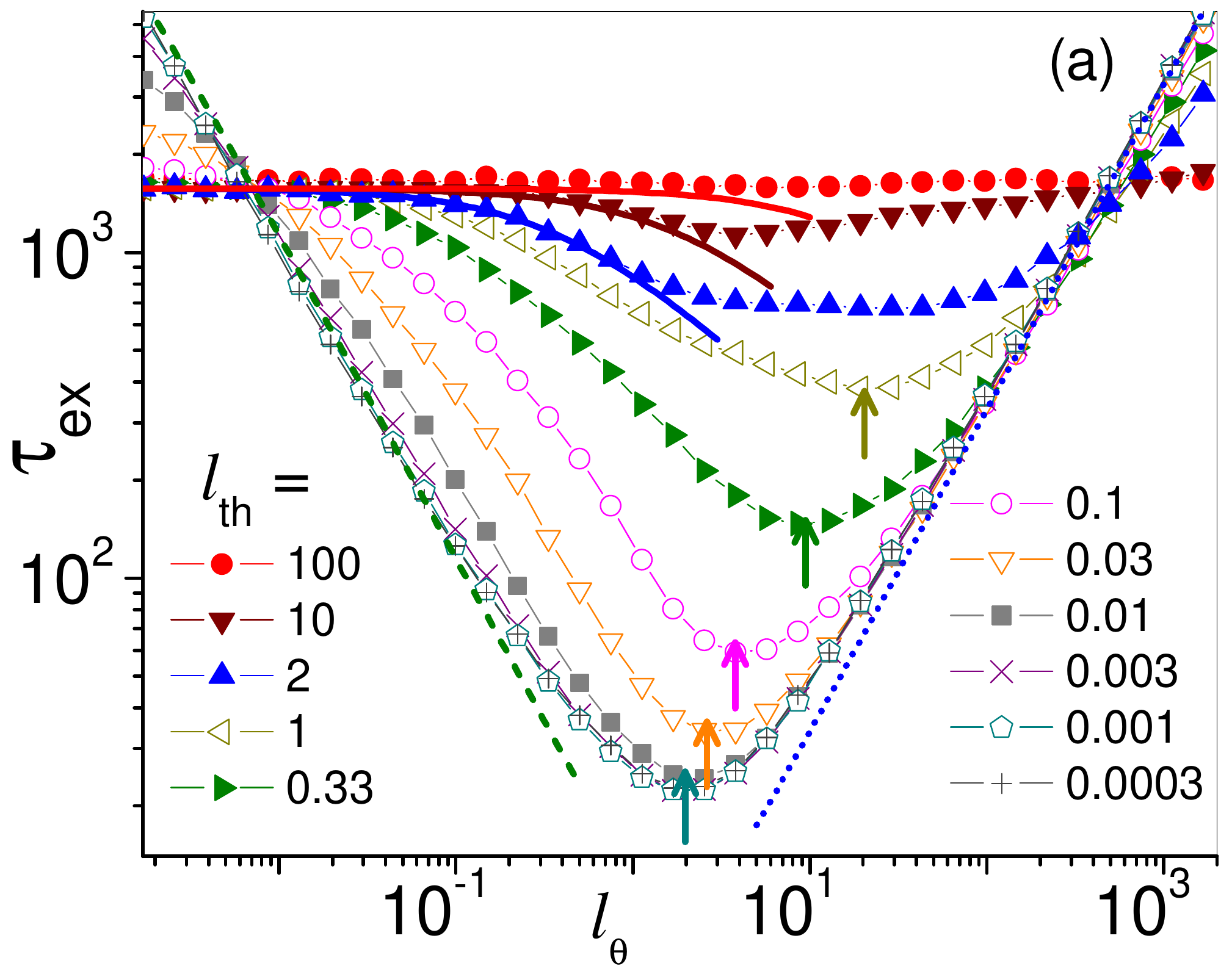}
\centering \includegraphics[width=7cm]{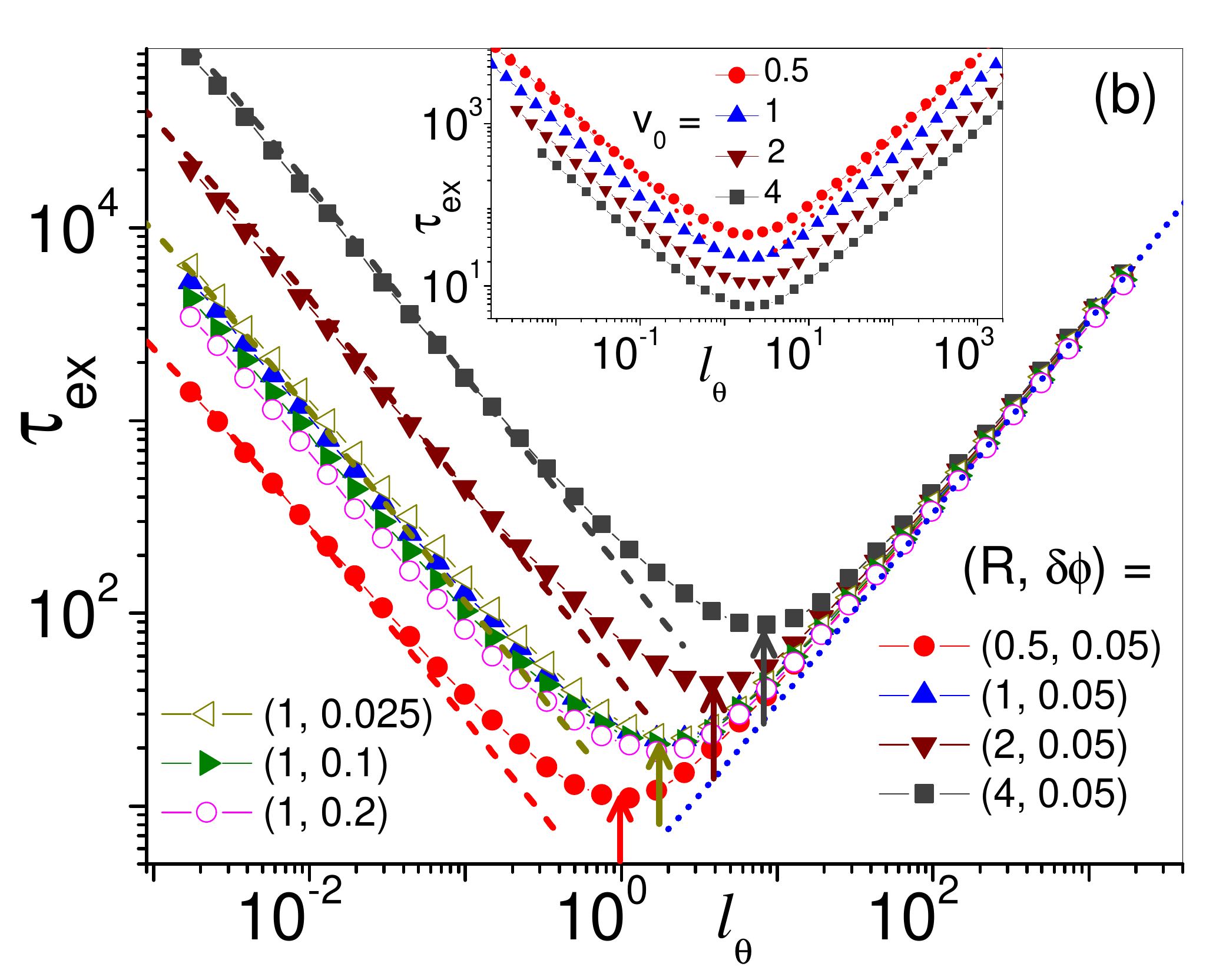}
\caption{ (Color online) (a) Mean exit time ($\tau_{ex}$) from a single-opening cavity as a function of self-propulsion length for different thermal lengths (shown in the legends). Analytical estimates for the positions of the minima depicted by vertical arrows. Estimates based on other analytical results represented by dashed lines Eq.~(\ref{exit-eq-3}), dotted line Eq.~(\ref{exit-eq-2}) and solid lines  Eq.~(\ref{un-1}).  (b) $\tau_{ex}$ {\it vs.} $l_{\theta}$ for different cavity and pore sizes as shown in the legends. Vertical arrows, dotted lines and dashed lines carry the same significance as panel (a). The inset of the panel (b) shows $\tau_{ex} \; vs. \; l_\theta $ for different self-propulsion velocity. Simulation parameters (unless mentioned otherwise ) %in the legends)
$v_{th} = 0.1,\; v_0=1.0, \;  R = 1, \; \delta \phi = 0.05, \; l_{th} = 0.001$.  \label{F1}}
\end{figure}

\section{Escape kinetics of self-propelled particles in the overdamped limit} 
Numerical simulation results of mean exit time ($\tau_{ex}$) for the various damping regimes, the self-propulsion length, and other parameters are presented in Fig.~(2-6). Our analysis mainly focuses on the interplay between self-propulsion and inertia in the escape kinetics of an active particle.  We first explore the behaviour of mean exit time in the overdamped limit. This limit is normally characterized by the fast viscous relaxation time compared to any other relevant time scales appearing in the escape kinetics. However, for diffusion in confined structures, such restriction is more conveniently expressed by comparing thermal length, $l_{th} = v_{th} \tau_\gamma$,  with the other pertinent length scales of the system. Where, thermal velocity is defined as, $v_{th} = \sqrt{k_B T/m}$.

In the overdamped limit, a passive particle's dynamics is governed only by its free space diffusion constant, $D_0 = k_BT/\gamma$. Further, the escape rate in this limit is directly proportional to the diffusion constant irrespective of the nature of the cavity and pore structures \cite{jcp2,Burada-review,holcman-review,Zwanzig}. Thus, the mean exit time becomes directly proportional to the damping constant $\gamma$. However,  overdamped self-propelled particles display different behaviours (see Fig.~2-3) depending upon values of parameters.

 As we have chosen the propulsion velocity, $v_0 = F_0/\gamma$, as an independent parameter, the escape kinetics of a highly damped active particle is insensitive to damping strength when self-propulsion effects dominate over the thermal translational diffusion. On the other hand, when both the self-propulsion and thermal motion contributions are comparable, the asymptotic features depend on the length of self-propulsion relative to the cavity radius and pore size.  We systematically explore escape dynamics out of single-opening, double-opening (with varying separations) and finally, multiple-opening cavities. 

\subsection{Cavity with a single exit window}
Figure 2(a) depicts variation of $\tau_{ex}$ with $l_\theta$ for different thermal lengths. Here, variations of thermal length amount to variations of damping strength with a fixed thermal velocity. It is apparent from simulation results that for large  damping, such that, $l_{th} $ is much smaller than the pore width $ \delta \phi R$, $\tau_{ex}$ is insensitive to the $l_{th}$  for the entire range of self-propulsion length as long as $D_0 << D_s$. This feature can be treated as an identifying mark for the overdamped limit.

 Simulation results of highly damped active particles [see Fig.~2(a, b)] show that $\tau_{ex}$ is inversely proportional to the self-propulsion length for $l_\theta << R$. The opposite limit, $l_\theta >> R$, witnesses a linear growth of escape time with $l_{\theta}$. In between these two limits, a minimum is observed in $\tau_{ex}$  versus  $l_\theta$.  Thus, it appears that different exit mechanisms are involved in the three regimes of $l_\theta$ or rotational dynamics. The distinct exit mechanisms can further be anticipated by looking at particle trajectories inside the cavities (see Fig.~1(c)).  For $l_\theta >> \{R, \; l_{th}\}$ and $l_{th} <<R$, our simulations results [see Fig.~1(d-e)] show that particles get accumulated and diffuse along 2D cavity circumference keeping $\vec{v_0}$ aligned to the position vector $ \vec{r}$. Such type of accumulation of overdamped active particles near boundaries or obstructions is a manifestation of their persistent motions.   This  feature triggers interesting phenomenologies reported in the earlier studies~\cite{VolpeP,jcp1,JP1,JP2,R8,R9,R10}.

{\it Slow rotational relaxation} --- To realize the escape mechanism out of a single-opening circular cavity for $l_\theta >> R$,  we first consider the situation where the particles  are initially positioned antipodal to the opening window and self-propulsion is anti-parallel to the pore direction. When the thermal velocity, $v_{th} < v_0$,  the particle diffuses on the cavity surface and $\vec{v_0}$ remains radially directed. Thus,  $\tau_{ex}$ here is basically time for rotational diffusion of an angle $\pi - \delta\phi/2 $ [see Eq.(\ref{A1}) in appendix A],  
\begin{eqnarray}
\tau_{ex} = \frac{(2\pi - \delta\phi)^2 l_\theta}{8v_0}.\label{exit-eq-1}
\end{eqnarray}
Let us consider different initial conditions, at $t=0$ particles are placed at the middle of the cavity and $\vec{v_0}$ orientation has uniform distribution within the range $[0-2\pi]$. Particles exit from the cavity can be considered as a two-step process. First, particles drift to the cavity wall with velocity $\sim \vec{v_0}$, then  rotational diffusion by an appropriate angle takes the particle  at the exit window. Further to note, a little fraction ($\delta\phi /2\pi$) of trajectories directly hit the opening window. Based on this escape mechanism, our analytic calculation gives [see Eqs.~(\ref{A2}-\ref{A4}) in appendix A],  
\begin{eqnarray}
\tau_{ex} = \frac{(2\pi - \delta \phi)^3 l_\theta}{24\pi v_0}+\frac{R}{v_0} \label{exit-eq-2}
\end{eqnarray}
Predictions based on this equation  well corroborate simulation results (shown in Fig.~2).  The second term in  Eq.~(\ref{exit-eq-2}) accounts for the approximate time to reach the wall starting from the cavity centre.   Further, the first term is much larger than the second one as $l_\theta >> R$.  Thus, $\tau_{ex}$  becomes almost insensitive to the position where the particles are initially placed. Moreover, for a very narrow opening Eq.~(\ref{exit-eq-2}) can be simplified as, $\tau_{ex} \sim \pi^2 l_\theta /3v_0$.  This result is consistent with previous studies \cite{R1,R2} on escape dynamics from circular cavities.

{\it Intermediate regime of rotational relaxation} --- With decreasing $l_\theta$ the mean exit time keeps decreasing until it reaches its minimum. The random search of the opening window becomes most effective when $\vec{v_0}$ changes its direction right after travelling a length equal to the cavity diameter. Thus, the minima in $\tau_{ex} \; vs. \; l_\theta$ plots are located at, $$(l_\theta)_{min} = 2R.$$
 The minima positions based on this estimate are marked by vertical arrows in Fig.~2(a, b). Further, minima positions are independent of initial conditions. As anticipated, simulation results show that the exit time at minima $\tau_{ex}^m$ is directly proportional to hitting frequency at the wall, thus,  $\tau_{ex}^m \propto R/v_0$. Further, both $\tau_{ex}^m$, as well as $(l_\theta)_{min}$, are insensitive to the width of opening windows. This can be understood based on the following reasoning. Inside the cavity, particles [see the trajectories in Fig.~1(c)] fly from one point to another wall point after a short stay at the cavity surface.  During their sojourn on the wall, they diffuse on the cavity surface over a length ($ \Delta $) much larger than the pore arch length. Consequently, when particles hit the wall within a distance $ \Delta $ from the pore, they most likely exit without flying to another point on the wall. This exit mechanism makes $\tau_{ex}$ insensitive to pore widths as long as $\Delta$ appreciably larger than $\delta \phi R$.

{\it Fast rotational relaxation} --- When $l_{\theta} <<  \delta \phi R$, as well as, $\tau_{\theta}$ is much shorter than any other relevant time scale of the system, it can be assumed that self-propellers exhibit almost uncorrelated diffusive motion inside the cavity. The effective diffusion constant of the active particle, $D_{eff}=D_0 + v_0^2/2D_\theta$, carries the contribution from both thermal motion as well as the self-propulsion. In this limit, the success rate in random search of the opening of the diffusive particle is directly proportional to its diffusivity. Following \cite{holcman-review} the exit time can be expressed as,   
\begin{eqnarray}
\tau_{ex} = \frac{R^2}{D_0+l_\theta v_0/2}\left[\ln \left(\frac{2\pi -\delta \phi}{\delta \phi}\right)+\chi\right],  \label{exit-eq-3}
\end{eqnarray}
where, $\chi$ is a constant and its value depends on the initial conditions. For exit from the antipodal point, the cavity center and uniformly distributed initial positions within the cavity, the approximate values of  $\chi$ are $ 2\ln2 $, $ \ln2 + 1/4$ and $ \ln2 + 1/8$, respectively~\cite{holcman-review}. As expected, the exit time is insensitive to the orientation of self-propulsion velocity at the starting point.  Further to note, for $D_0 \rightarrow 0$, Eq.~(\ref{exit-eq-3}) collaborates finding in Refs.~\cite{R1,R2}. 

 Simulation results presented in Fig.~2 well accord with Eq.~(\ref{exit-eq-3}). It appears from both simulation data, and analytic estimations, $\tau_{ex}$ has $\gamma$ dependence through thermal diffusion constant $D_0$. When self-propulsion dominates over thermal diffusion, $D_{eff} \approx v_0l_\theta/2$, the exit time $\tau_{ex}$ becomes independent of the damping strength. Thus, the overdamped limit for fast rotation diffusion limit can be recognized by the inverse relation between $\tau_{ex}$  and  $l_{\theta}$. Deviation from this relation in the fast rotational diffusion limit could be considered as inertial impact.
 
 We conclude our discussion on escape kinetics in the diffusive limit with a remark on the regime of $l_{\theta}$, $ R \gg l_{\theta} \geq \delta \phi R$. Under this restriction, active particles still exhibit diffusive motion inside the cavity. Nevertheless, after hitting the boundary wall, the particles diffuse on the cavity surface over a length comparable to the pore width. This diffusion on the surface leads to modification of the effective pore width. Thus, deviation from Eq.~(\ref{exit-eq-3}) gets noticeable as soon as self-propulsion length grows larger than the pore width.

\begin{figure}[tp]
\centering \includegraphics[width=7cm]{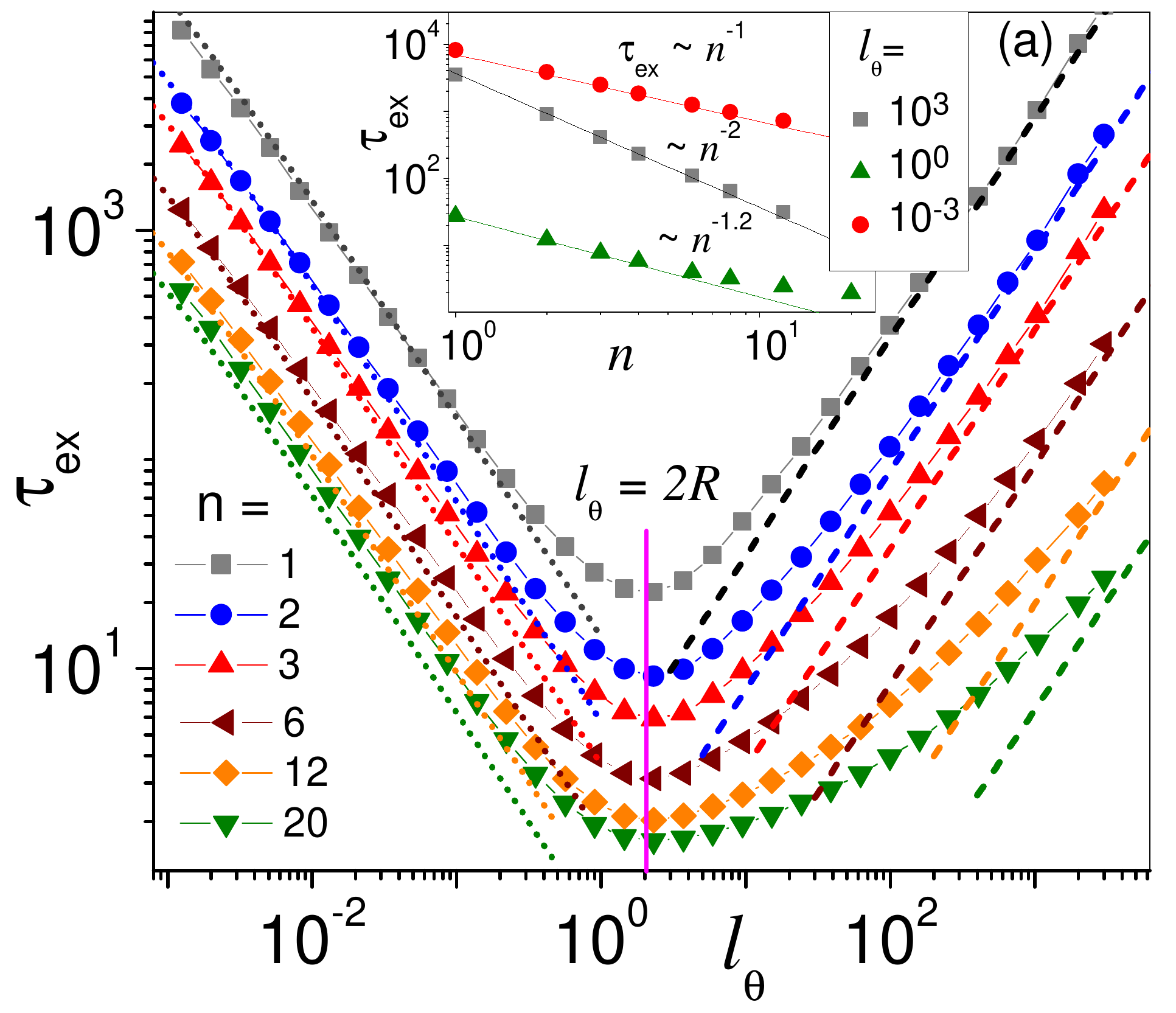}
\centering \includegraphics[width=7cm]{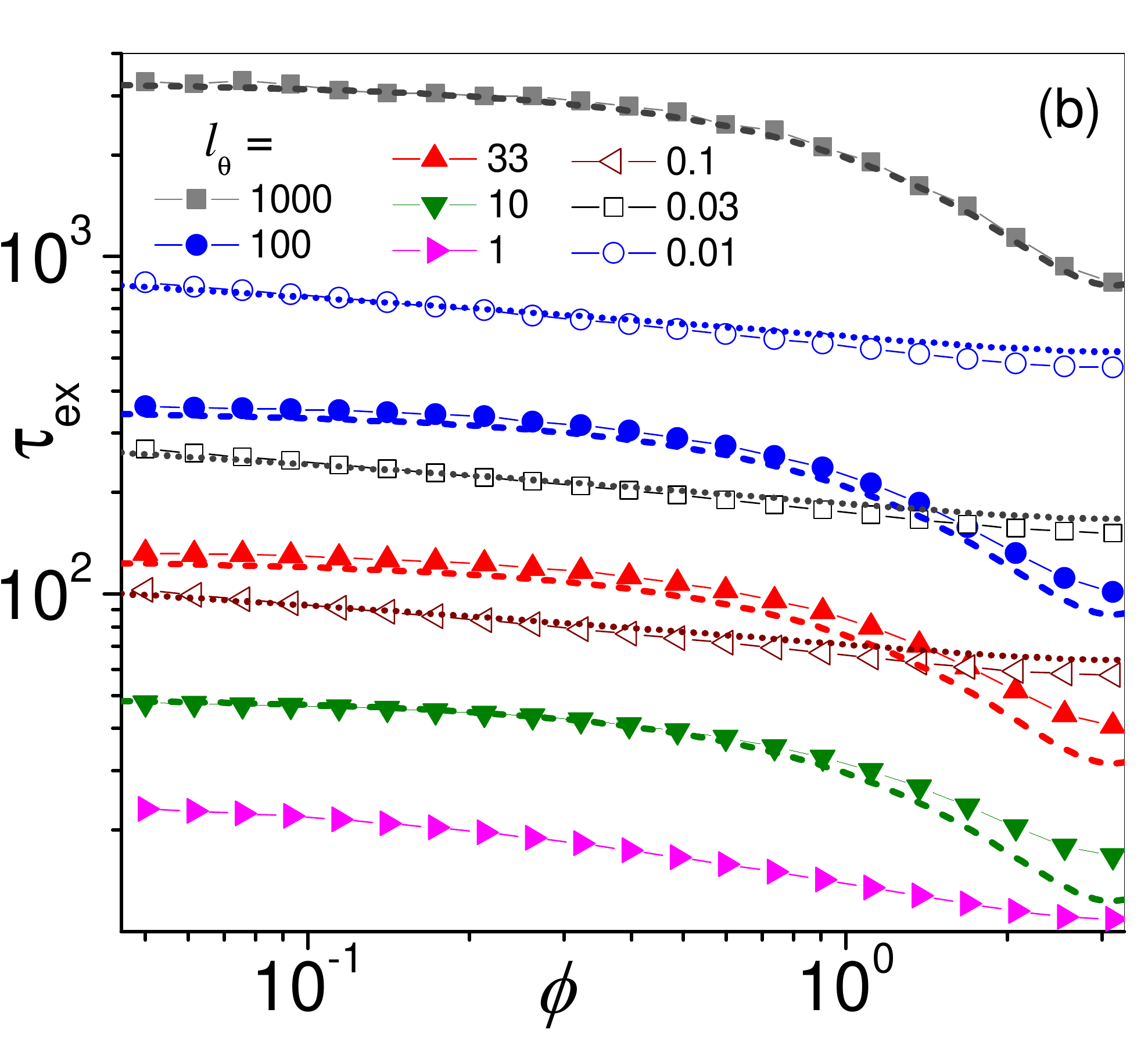}
\caption{ (Color online) (a) Mean exit time ($\tau_{ex}$) as a function of $l_\theta$ for different number of exit windows.  The pores are of the same size and they are equally spaced. The dashed and dotted curves
represent the corresponding analytical predictions of Eqs.~(\ref{npore-1}) and (\ref{npore-2}), respectively. Simulation parameters: $v_{th} = 0.1,\; v_0=1.0, \;  R = 1, \; \delta \phi = 0.025, \; l_{th} = 0.001$. Inset depicts variation of $\tau_{ex}$ with pore number $n$ for $l_{\theta} >> R$, $l_{\theta} \sim R$ and $l_{\theta} << R$. For these three regimes of rotational relaxations, simulation data are well fitted with the power laws: $\tau_{ex} \sim n^{-2}$, $ n^{-1.25}$ and $ n^{-1}$, respectively.   (b) Mean exit time from a two-opening cavity as a function of separating angle $\phi$ between two pore centers for different $l_\theta$ (see legends). The dashed and dotted curves are our analytical predictions from Eqs.~(\ref{2pore-1}) and (\ref{2pore-2}), respectively. Other parameters are same as (a), but $\delta \phi = 0.05$.       \label{F1}}
\end{figure}

\subsection{Circular cavity with double exit windows} We now study the impacts of an additional exit window on the narrow escape dynamics. Figure 1(b) shows that two pore centres are separated by an angle $\phi$. Thus,  the chordal distance between pore centers is, $\lambda = 2R\sin(\phi/2)$. Figure~3(b) shows the variation of $\tau_{ex}$ as a function of $\phi$ for different persistence lengths of self-propulsion.

As discussed  earlier, for a very large self-propulsion length, $l_\theta >> R$, the exit process rests on diffusion on the cavity surface as long as the $F_0$ dominates over the inherent thermal fluctuations. If the starting point of a particle trajectory is the cavity center, it hits the wall within the time $\sim R/v_0$ and the subsequent diffusion on the wall surface takes the particle to the exit window. Further, active particle dynamics on the cavity wall is solely governed by its rotational diffusion. To estimate $\tau_{ex}$ analytically, particle trajectories are divided into the following three categories  based on their hitting points on the wall. (i) As little as  $\delta\phi/\pi$ fraction of particles  directly hit exit windows and they escape without diffusing on the wall. (ii) $(\phi-\delta\phi)/2\pi$ fraction of particles arrive on the wall within the polar angle $\delta\phi/2 \; {\rm to} \; \phi - \delta\phi/2$ [see Fig.~1(b)], and they need to diffuse over the angle $ 0 \; {\rm to} \; (\phi - \delta\phi)/2$ to arrive the nearest exit point. (iii) The remaining particles find the opening window by diffusing over the angle, $ 0 \; {\rm to} \; (2\pi - \phi - \delta\phi)/2$, right after arriving at the cavity surface.  Based on this exit mechanism, we derive an analytic expression for $\tau_{ex}$ [see Eqs.~(\ref{A5}-\ref{A6}) in appendix A],    
\begin{eqnarray}
\tau_{ex} = \frac{l_\theta}{24\pi v_0}\left[ (\phi - \delta \phi)^3+(2\pi-\phi - \delta \phi)^3\right]+\frac{R}{v_0} \label{2pore-1}
\end{eqnarray}
This equation well corroborates simulation results presented in Fig.~3. However, it loses its validity when the self-propulsion persistence length is reduced to the order of cavity dimension. In this limit, as long as the diffusion length $\Delta$ on the wall is larger than the pore separations, $\tau_{ex}$ remains almost insensitive to $\phi$. Variations of $\tau_{ex}$ with $\phi$ become noticeable when $\lambda$ goes beyond the diffusion length $\Delta$.

 On further reduction of $l_\theta$, particles dynamics inside the cavity become completely diffusive. Recall that this regime is characterized by the restriction,  $l_{\theta} \ll \delta \phi R$.  For well separated opening windows, the exit process through one window is not affected by the other adjacent pores.  Thus, total exit rate can be expressed as a sum of the exit rate through individual windows. This leads to, $\tau_{ex} =  \tilde{\tau}_{ex}/2$.  Where,  $\tilde{\tau}_{ex}$ is the mean exit time through one opening when the other windows are closed. On the other hand, when the exit points are not well separated, a logarithmic dependence of $\tau_{ex}$ on the chordal distance between pore centers is observed~\cite{holcman-review} 
\begin{eqnarray}
\tau_{ex} = \frac{2R^2}{2D_0+l_\theta v_0}\left[\ln \left(\frac{2\pi R}{\lambda}\right)+\tilde{\chi}\right]. \label{2pore-2}
\end{eqnarray}
where, $\tilde{\chi}$ is a constant which depends on the pore size as well as initial conditions. For very narrow exit windows and when the particle is placed at the center of the cavity at the starting point, $\tilde{\chi}$ approaches to $\ln [2(2\pi-\delta \phi)/\delta \phi]+O(1)$. We compare estimations based on the Eq.~(\ref{2pore-2}) with our simulation results. They fairly corroborate each other.    

\subsection{Circular cavity with multiple exit windows}
It is known that the mean escape time of passive Brownian particles from a cavity with well-separated exit pores is inversely proportional to the number of pores \cite{holcman-review,JPCC-1}. However, our simulation results show that for active particles, $\tau_{ex} \sim 1/n^{\alpha}$. The exponent $\alpha$ varies in the range from 1 to 2  depending  upon the persistence length of self-propulsion. For $l_{\theta} \gg R$, $\alpha$ assumes its maximum value $2$. In the opposite limit  $l_{\theta} \ll R$, $\alpha = 1$.

Recalling the exit mechanisms of active particles for very large $l_{\theta}$, in the present case, the particle trajectories can be divided into two categories. Some trajectories directly hit the opening, thus taking approximately $R/v_0$ time to exit the cavity. Another type of trajectories require rotational diffusion over the angle  $0 \; {\rm to} \; \pi/n-\delta \phi$  after hitting the wall, to reach the nearby exit window. Based on the same line of reasoning as Eqs.~(\ref{2pore-1}, \ref{A2})  we deduce the following expression for the mean exit time,   
\begin{eqnarray}
\tau_{ex} = \frac{l_\theta}{24\pi v_0 n^2}\left(2\pi - n\delta \phi\right)^3+\frac{R}{v_0} \label{npore-1}
\end{eqnarray}
Here, the first term contributes in exit time much larger than the second one as $l_{\theta} >> R$. Further, for a very narrow opening window, simplification of Eq.~(\ref{npore-1}) produces, $\tau_{ex}\sim \pi^2 l_{\theta}/3v_0 n^2$. Thus, this result supports well fitted power law $\tau_{ex} \sim 1/n^2$ in the inset of Fig.~3(a)  for large $l_\theta$. Moreover, assessments %estimations 
based on Eq.~(\ref{npore-1}) represented with dashed lines in Fig.~3(a) accord with simulations.

Simulation results show that in the intermediate regime of $l_\theta$, the exponent $\alpha$ is still larger than 1. However, the position of the minimum in $\tau_{ex} \;  vs. \; l_\theta$ is insensitive to the number of pore $n$. Moreover, $\tau_{ex}^m$ becomes independent of  $n$   as soon as the spacing between pores approaches the diffusion length on the wall.

 In the very fast rotational dynamics of the active particle, the self-propulsion adds up to uncorrelated thermal fluctuations.  Thus,  following\cite{holcman-review} one can find the expression of mean exit time as,  
\begin{eqnarray}
\tau_{ex} = \frac{2R^2}{n(2D_0+l_\theta v_0)}\left[\ln \left(\frac{2\pi R}{\delta \phi}\right)+\frac{n}{8}-\ln \frac{2}{n}\right]. \label{npore-2}
\end{eqnarray}
Simplification of this equation for very narrow and well separated pores yields, $\tau_{ex} \sim \tilde{\tau}_{ex}/n$. As mentioned earlier,  $\tilde{\tau}_{ex}$ is the mean exit time when only one opening window is there. Thus, when pores are well separated, the exit through one pore is not affected by the presence of other windows.     

\begin{figure}[tp]
\centering \includegraphics[width=6.5cm]{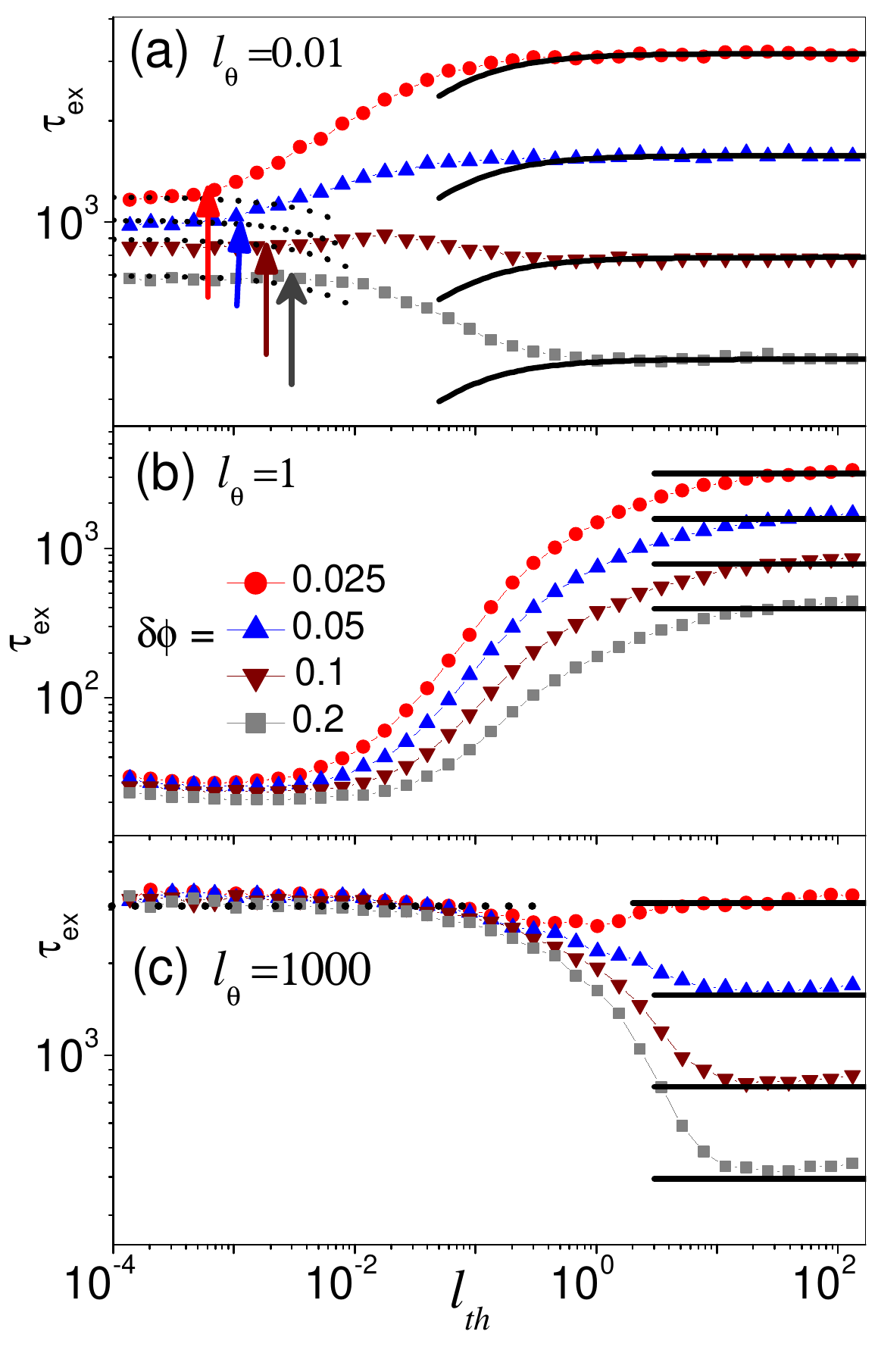}
\centering \includegraphics[width=6.5cm]{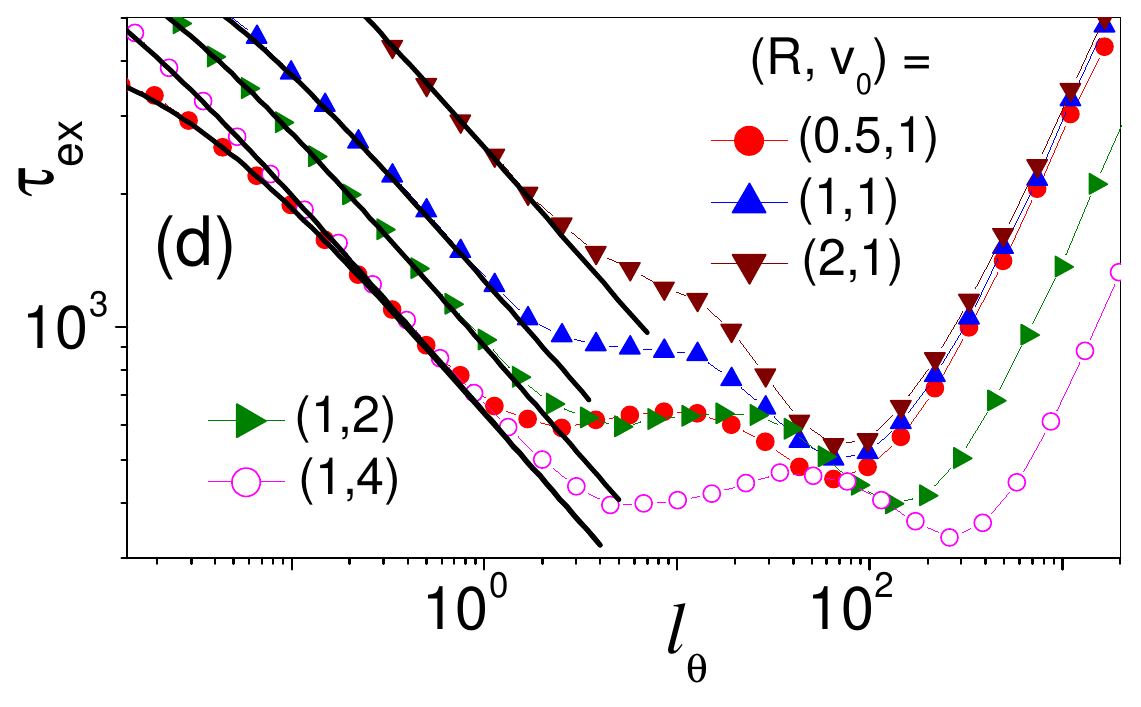}
\caption{ (Color online)(a-c) Mean exit time ($\tau_{ex}$) from a circular cavity with a single opening as a function of the thermal length $l_{th}$ for different pore sizes (shown in the legends). The thermal length is varied by the changing damping constant $\gamma$ where as the thermal velocity is kept fix.  The panels represent three regimes of self-propulsion length: (a) $l_\theta << R$, 
(b) $l_\theta \sim R$, and (c) $l_\theta >> R$. Analytical estimates for the asymptotic regimes depicted by black dotted [Eq.~(\ref{exit-eq-3}) for (a), Eq.~(\ref{exit-eq-2}) for (c)]  and
solid lines [Eq.~(\ref{un-1})]. Other simulation parameters (unless mentioned in the legends) $v_{th} = 0.1,\; v_0=1.0, \;  R = 1, \; \delta \phi = 0.05$. (d) Mean exit time versus self-propulsion length for different self-propulsion velocities and cavity sizes. Other parameters are $v_{th} = 0.017,\; l_{th} = 0.6 \; {\rm and } \; \delta \phi = 0.05$.    \label{F1}}
\end{figure}

\section{Escape kinetics of self-propelled particles in the underdamped limit} 
The simulation results presented in Fig.~2(a) and Fig.~4) show the impact of inertia in the escape kinetics from a single-opening circular cavity.  Figure 4 depicts  $\tau_{ex} \; vs. \; l_{th}$, in the distinct regimes of self-propulsion lengths $l_\theta$. We mention the fact that variations of thermal length are made by changing $\gamma$ for an unaltered thermal velocity.
%Note that here variations of thermal length are made by changing $\gamma$ for a given thermal velocity.
All variations in Fig.~4 display the common asymptotic features. In both limits, $l_{th}\rightarrow 0$  and $\rightarrow  \infty$ , the escape rate from the cavity is independent of thermal length (viz. damping strength of the medium). To figure out the parameter regimes where inertial impact cannot be ignored, the results of the overdamped limit presented in the previous section can be used to set references. Features in escape kinetics are systematically presented for single, double and multiple openings cavities.  Further, the impacts of inertia in the three regimes of rotational relaxation time have received due deliberation. %analyzed separately.  

\subsection{Cavity with a single exit window}
{\it{Fast rotational relaxation}} -- As discussed earlier, when $l_\theta << R$ and $\tau_\theta $ is shorter than any relevant time scale of the system, 
the effect of self-propulsion amounts to the augmentation %enhancement 
of the effective temperature of the system. Thus, self-propellers can be pondered %treated
as passive particles diffusing in a viscous medium at the effective temperature,   
\begin{eqnarray}
T_{eff} = \frac{\gamma}{k_B}\left(D_0+\frac{l_\theta v_0}{2}\right). \label{Teff}
\end{eqnarray}
In the very low damping regimes, when the particle travels a length within the viscous relaxation time is larger than the cavity diameter, the particle moves back and forth inside the cavity [see Fig.~1(c)] and occasionally exits through the narrow opening with a probability $\delta \phi /2\pi$. Furthermore, for Maxwellian distribution of velocity, the particle drifts to the wall with the average velocity, $\sqrt{2k_BT_{eff}/m\pi}$. Therefore, an apprehension of collision frequency and their success probability steer the mean exit time as ~\cite{JPCC-1}:
\begin{eqnarray}
\tau_{ex} =  \frac{2\pi R}{\delta \phi} \sqrt{\frac{\pi m}{\left(2k_B T + m v_0 v_{th}l_\theta/ l_{th} \right)}}. \label{un-1}
\end{eqnarray}
We note that  
the validity of this equation is strictly restricted  for effective thermal persistence length, $\tau_\gamma\sqrt{k_B T_{eff}/m} > 2R$ in addition to $l_\theta <<R$. The appraisals 
based on the Eq.~(\ref{un-1})  [shown by solid black lines in Fig.~4(a)] fairly accord with simulation results.     

Recall that when escape kinetics is free from the inertial impact, the mean escape time is inversely proportional to the effective diffusion constant. Thus, one can identify the limit until inertia effects in the exit process  make their presence felt by observing deviation from the following empirical relation \cite{holcman-review},        
\begin{eqnarray}
\tau_{ex} = \frac{\Lambda}{k_BT/\gamma+v_0l_\theta/2}. \label{un-2}
\end{eqnarray}
Where $\Lambda$ is a constant which depends on the cavity and pore sizes in addition to the initial conditions. For a very narrow pore size, $\delta \phi /2\pi \rightarrow 0$, the expression of $\Lambda$ can be obtained from Eq.~(\ref{exit-eq-3}). For an arbitrary size and design %shape  
 of the opening window, the numerical value of $\Lambda$ can be extracted from simulation data using least square fitting. The critical thermal length $l_{th}^c$ represents the onset of an overdamped regime depicted by vertical arrows in Fig.~4(a). As expected, the inception of an %the onsets of the 
overdamped regime moves to smaller values of thermal length with decreasing pore size. It is apparent from the simulation results that the impact of inertia can safely be ignored as long as the effective thermal persistence length is much shorter than the shortest length scale involved in the escape kinetics. This requires the restriction, $\tau_\gamma \sqrt{k_BT_{eff}/m} < \delta \phi R$ for the problem at hand. Based on this requisition, %restriction,
$l_{th}^c$ can be estimated as,  
$$ l_{th} ^c = \frac{1}{2}\left(\sqrt{\frac{v_0^2 l_\theta^2}{4v_{th}^2}+4\delta \phi^2 R^2}-\frac{v_0 l_{\theta}}{2v_{th}}\right) $$
The predicted threshold value is close to the observed one marked by vertical arrows in Fig.~4(a).

{\it Slow rotational relaxation} --- On the other hand, when self-propulsion length is much larger than the cavity size, $l_\theta >> R $, the impact of self-propulsion can no longer be accounted for through the effective temperature of the system. Simulation results show that for the vanishingly small damping (such that $l_{th}>>R$), irrespective of the values of self-propulsion velocity and rotational relaxation time, particles randomly oscillates back and forth inside the cavity until they exit by hitting at the opening window. The thermal translational motion largely oversees %controls
the motion inside the cavity and the motion weakly turns on %depends on 
the self-propulsion. The mean exit times in this limit is given by~\cite{JPCC-1},  
\begin{eqnarray}
\tau_{ex} =  \frac{2\pi R}{\delta \phi} \sqrt{\frac{\pi m}{2k_B T }} \label{un-3}
\end{eqnarray}  
The theoretical prediction well corroborates simulation results for $l_{th} >> R$. With decreasing thermal length, the persistent self-propulsion motion takes control over particle dynamics inside the cavity. Here, the overdamped limit is identified by $\gamma$ independent mean exit time given by the Eq.~(\ref{exit-eq-2}). Thus, the inertial impact free regimes can easily be perceived 
from $\tau_{ex} \; vs. \; l_{th} $, presented in Fig.~4(c).  

\begin{figure}[tp]
\centering \includegraphics[width=7cm]{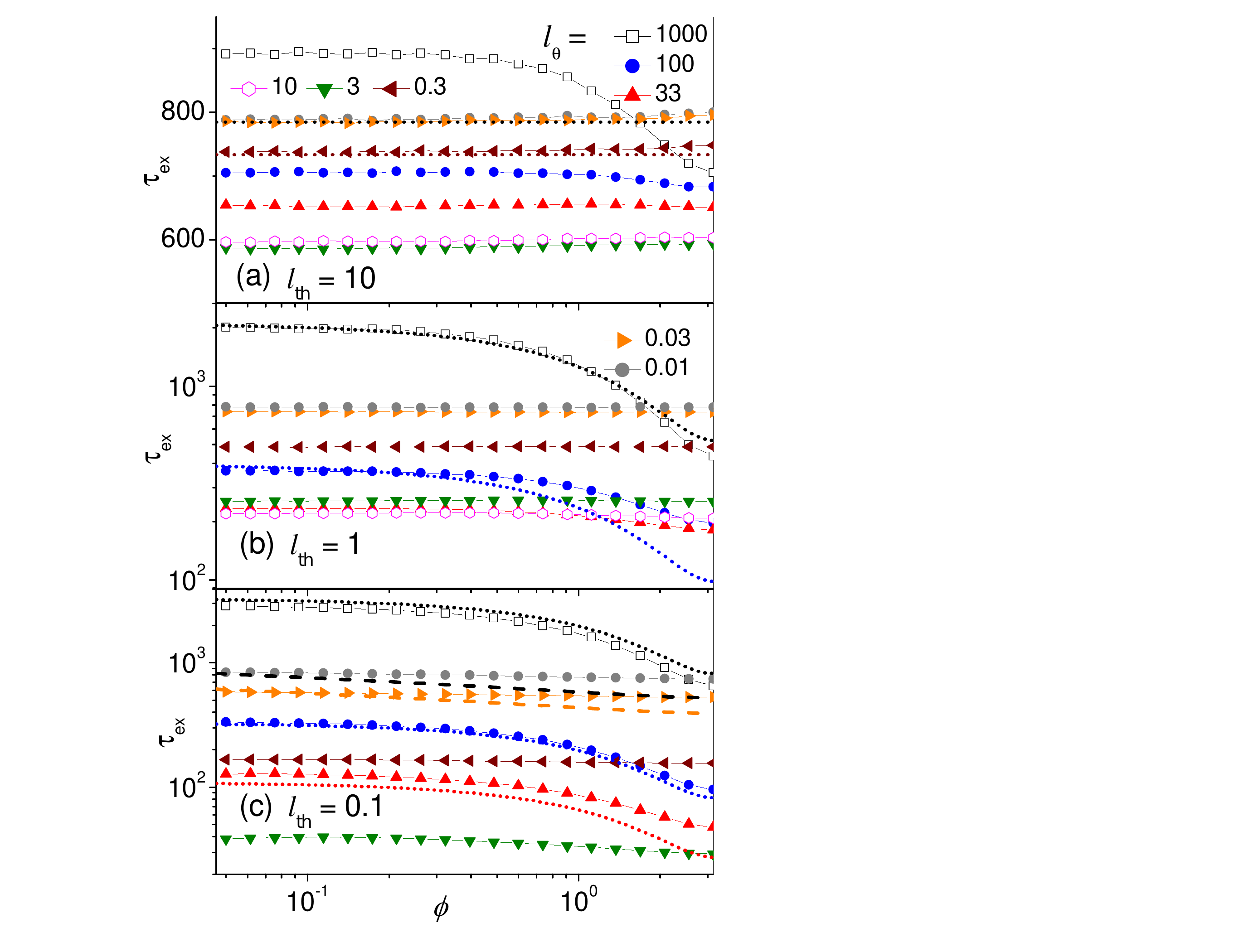}
%\centering \includegraphics[width=7cm]{Fig4d.pdf}
\caption{ (Color online) Mean exit time ($\tau_{ex}$) from a double-opening circular cavity as a function of separating angle $\phi$ between two pore centers for different $l_\theta$ (see legends).  Three panels (a), (b) and (c), corresponds to the three regimes of thermal lengths, $l_{th} >> R$, $l_{th} \sim R$  and $l_{th} \sim R\delta \phi$.  The dotted curves in (a) prediction based on Eq.~(\ref{un-1}). The dotted [ in panels (b, c)] and dashed curves [in the panel (c)] are our analytical predictions from Eqs.~(\ref{2pore-1}) and (\ref{2pore-2}), respectively. Simulation parameters (unless mentioned in the legends): $v_{th} = 0.1,\;  \delta \phi = 0.05, \; R = 1.0, \; {\rm and } \; v_0  = 1.0$.    \label{F1}}
\end{figure}

 {\it Intermediate regime of rotational relaxation} -- For a proportionate scale of the cavity size and  self-propulsion length, the exit time graces minimum in the overdamped regime as reported in the foregoing section. %When the self-propulsion length is comparable with the cavity size,the exit time becomes minimum in the overdamped regime as analyzed in the previous section. 
 However, in the very low damping regime, the escape kinetics for $l_\theta \sim R$  evinces kindred hallmarks %exhibits similar features 
 like other two limits of rotational dynamics, $l_\theta >> R \; {\rm and}  \; l_\theta << R $. The asymptotes here too can approximately be determined using Eq.~(\ref{un-3}). Moreover, escape kinetics carries inertial impact until, $l_{th} > \delta \phi R$. 

Simulation results in Fig.~2(a) show that in the underdamped limit, the positions of the minima in $\tau_{ex} \; vs. \; l_\theta $  have been shifted %are being shifted 
to the longer self-propulsion length. The location of the minima hinges on both the damping strength as well as the self-propulsion velocity.  %Moreover, the position of minima depends on both the damping strength as well as the self-propulsion velocity. 
This trait is attributed to the emergence of another length scale, $l_F = v_0 \tau_\gamma$. Due to its impact, the escape rate becomes maximum when the self-propulsion motion persists over a length $ 2l_F $ above the cavity diameter. The position of the minima estimated based on this reasoning, 
\begin{eqnarray}
(l_\theta)_{min} = 2R+\frac{2v_0 l_{th}}{v_{th}}, \label{minima-1}
\end{eqnarray}
represented by vertical arrows in Fig.~2(a). They are fairly consistent with simulation results.          
  
For $l_{th} \gtrsim R$, an additional minimum emerges in $\tau_{ex} \; vs. \; l_\theta$  around point of transition from the diffusive to ballistic motion of self-propulsion. However, in most of the cases, it appears very subtle to be noticed or merges with the minimum described in Eq.~(\ref{minima-1}).  As evident in %As it is witnessed in 
Fig.~2(a) and Fig.~4(d), the minimum at low $l_\theta$ becomes detectable
 %gets noticeable
when self-propulsion contributes to the effective temperature larger than %the intrinsic thermal part 
$k_BT$ around the point of transition $l_\theta \sim R$. Parameters are chosen accordingly in Fig.~4(d) to get well separated two prominent minima. The appearance of the left minimum in $\tau_{ex} \; vs. \; l_\theta$ can thus be justified as follows. When $l_\theta < R$ and $l_{th} \gtrsim R$, the self-propulsion motion acts as uncorrelated random fluctuation, which adds up to the inherent thermal fluctuations leading to enhancement of the system effective temperature. Thus, the mean exit time decays with self-propulsion according to the Eq.~(\ref{un-1}). Once the $l_\theta$ length is comparable to the cavity size, the contribution of self-propulsion in the effective temperature subsides.  %As soon as the $l_\theta$ length becomes comparable to the cavity size, the contribution of self-propulsion in the effective temperature starts decreasing. 
As soon as $l_\theta$  grows much longer than cavity diameter,  the self-propulsion motion, instead of contributing to effective temperature, affects escape kinetics by different mechanisms.  Such chop and change %This switching 
between two self-propulsion regimes escorts %leads
to the left minimum.

\begin{figure}[tp]
\centering \includegraphics[width=7cm]{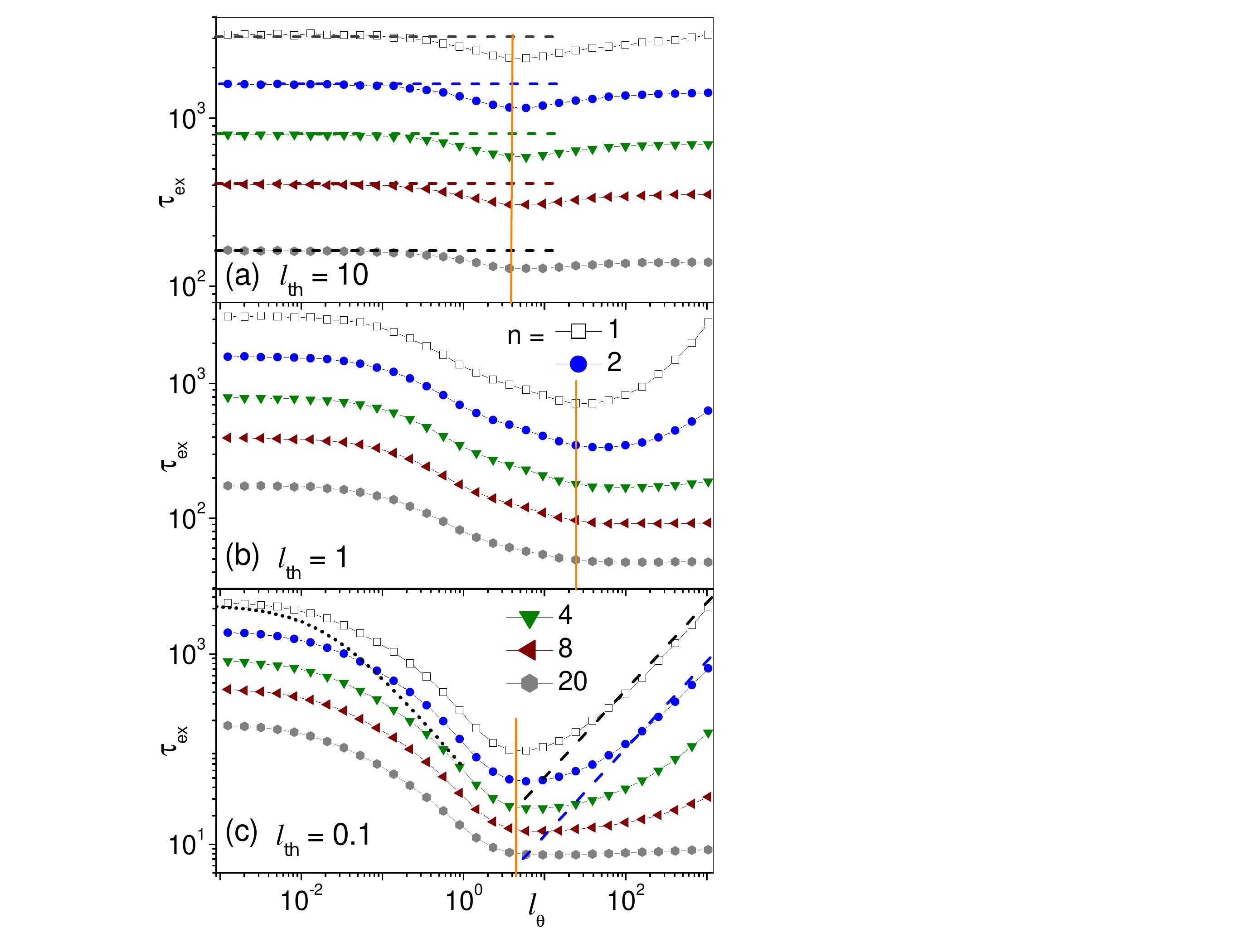}
%\centering \includegraphics[width=7cm]{Fig4d.pdf}
\caption{ (Color online) Mean exit time ($\tau_{ex}$) from a cavity as a function of self-propulsion length $l_\theta$ for different number of opening windows $n$ (shown in the legends). Three panels (a), (b) and (c), corresponds to the three regimes of thermal lengths, $l_{th} >> R$, $l_{th} \sim R$  and $l_{th} \sim R\delta \phi$. The dashed curves in (a) are estimations based on Eq.~(\ref{un-1}). The dashed and dotted curves in (c) are our analytical predictions from Eqs.~(\ref{npore-1}) and (\ref{npore-2}), respectively. Solid vertical lines in all three panels mark the position of minimum. Simulation parameters (unless mentioned in the legends): $v_{th} = 0.1, \; \delta \phi = 0.025, \; R = 1.0, \; {\rm and } \; v_0  = 1.0 $.    \label{F1}}
\end{figure}

\subsection{Circular cavity with multiple exit windows}
Our simulation results reported in Fig.~5-6 present effects of multiple exit windows in the escape kinetics of an underdamped active particle. Three panels in Fig.~5 show variations of $\tau_{ex}$ with separation angle $\phi$ between two pores centres for various relative amplitudes of $l_\theta$ and $l_{th}$ with a fixed cavity radius. Here, our analysis considers three values of $l_{th}$, that correspond to the limits, $l_{th} \gg R$, $l_{th} \sim R$ and $l_{th} \sim R\delta \phi$, respectively.  

For, $l_{th} \gg R$ and  $l_{\theta} \ll R$,  based on the arguments of the single-opening cavity, we expect $\tau_{ex}$ to be insensitive to the separation between two pores. Thus, $\tau_{ex}$ can be estimated using Eq.~(\ref{un-1}), however, doubling the pore width.  Our expectations are supported by simulation results presented in Fig.~5(a). With gradually increasing self-propulsion length, the exit time remains independent of $\phi$ as long as $l_\theta \leq l_{th}$. As the self-propulsion length grows beyond this limit, the escape processes become sensitive to the pore separation angle $\phi$ [see Fig.~5(a)]. This feature could be understood assuming qualitative similarity of the exit mechanism with highly damped active particles. For $l_\theta \gg l_{th}$, most likely, due to the enhanced probability of accumulation and diffusion on walls, the narrow escape process gets expedited with increasing separation between pores.   
 
Figure 5(b) shows that when the thermal length is comparable to cavity radius, the mean exit time is almost independent of the separation between two pores until $l_\theta < l_{th}$. However, in the opposite limit of self-propulsion lengths $l_\theta \gg l_{th}$, particles accumulate on the cavity wall and the rotational diffusion controls the particle's escape dynamics. Thus, $\tau_{ex}$ closely follows the exit mechanism discussed in the context of Eq.~(\ref{2pore-1}).  Simulation results here are fitted well  Eq.~(\ref{2pore-1}) with an adjustable pre-factor. 

For,  $l_{th} \sim  \delta \phi R$, simulation results show that the escape kinetics shares very similar features with overdamped regimes. However, the mean exit rate is not free from the impact of inertia as diffusion across the pore gets correlated due to inertial memory. This effect can be accounted for by considering a $\gamma$-dependent effective pore widths. Thus, simulation data could be fitted with  overdamped  Eqs.~(\ref{2pore-1}-\ref{2pore-2}), however, one needs $\gamma$ dependent adjustable pore-width.

 Figure~6 shows $\tau_{ex} \; vs. \; l_{\theta}$ with varying number of exit windows. The opening windows are uniformly distributed over the crater %cavity 
 wall as mentioned in the context of overdamped particles.  In contrast to the overdamped limit, our simulation results show that  $\tau_{ex} \sim 1/n$ for $l_{th} >> R$ and irrespective of the values of self-propulsion length.  This implies that the exit process through a pore is not affected by the presence of other opening windows no matter how close they are. Following Eqs.~(\ref{un-1}, \ref{un-3}),  one can estimate the mean exit time from a circular cavity with $n$ pores as~\cite{JPCC-1},   
\begin{eqnarray}
\tau_{ex} =  \frac{1}{n}\frac{2\pi R}{\delta \phi} \sqrt{\frac{\pi m}{2k_B T_{eff} }} \label{un-4}
\end{eqnarray} 
Note that this equation is strictly restricted for $l_\theta << R\delta \phi$ in addition to very slow viscous relaxation. Further,  as discussed earlier, the self-propulsion motion contributes through effective temperature given by the Eq.~(\ref{Teff}). This analytic estimations fairly corroborate the simulation results, as depicted in Figs.~5-6. 

  For all the three values of thermal lengths, corresponding to the panels (a, b, c) in Fig.~6,  $\tau_{ex} \; vs. \; l_{\theta}$ exhibits a minimum. The position of the minimum is insensitive to the number of exit windows the cavity possesses. For $l_{th} \sim  \delta \phi R$ and $l_{th} \sim  R$, the minimum can be located using Eq.~(\ref{minima-1}). However, for $l_{th} \gg  R$, the minimum appears around the point of transition of self-propulsion motion from the diffusive to the ballistic regime.

\section{Conclusions} 
Present study offers a structured appraisal of the escape kinetics of an active particle from a circular cavity with multiple pores in the different dominion of self-propulsion parameters, damping strengths and inbred thermal diffusion.  We encode the self-propulsion effects on the dynamics by considering the active particle is of the Janus kind. However, this modelling would apply to other microswimmers of biological nature as well. Both simulation and analytic results show that the relative dominance of four length scales, namely thermal length $l_{th}$, self-propulsion persistence length $l_{\theta}$, cavity diameter $2R$ and width of exit windows $R \delta \phi$, controls the escape through narrow openings. The pivotal aspects emerging in the different commands of these length scales can be summarized as follows:  
 
 (i) Inertial impact-free regimes in the escape kinetics are characterized by inequality, $l_{th} << R \delta \phi$. It is manifested by damping strength independent exit rate when the self-propulsion motion largely governs active particle dynamics. However, in the weak self-propulsion limit, the narrow escape rate is inverse to the damping strength.

 (ii) In the overdamped limit,  active particles diffuse on walls for an overwhelmingly large fraction of their lifetime inside the cavity because of very slow rotational dynamics. Only rotational diffusion by an appropriate angle can take the active particle to the exit window. Thus, the narrow escape rate here is directly proportional to the rotational diffusion constant $D_{\theta}$ and insensitive to the cavity size, thermal translational diffusion, pore size and self-propulsion velocity. However, this mechanism does not work as soon as bouncing effects due to inertial footprints get pronounced.    

 (iii) When active particles' rotational relaxation is much faster than other relevant time scales, self-propulsion motion becomes as incoherent as the intrinsic thermal fluctuations. Thus, fluctuations from two sources can be added up to account for the effective diffusion and temperature.  In consequence, the self-propeller's escape mechanism gets similar to the passive Brownian particles with a modified strength of thermal noise.
  
(iv) For highly damped active particles, the mean exit time versus self-propulsion length exhibits a minimum when self-propulsion length approaches the cavity diameter. This sort of resonant-like phenomenon \cite{Doering} attributes to the synchronization of rotational relaxation and swimming time over the cavity 
diameter. This effect is insensitive to thermal noise, the pore number, their widths and separation of pores (in cases of double opening). Further, this type of synchronization effect prevails in the underdamped regimes. However, the position of minima in $\tau_{ex}\;vs.\; l_{\theta}$ depends on the damping strength. The synchronisation gets blur for $l_{th} >> R$.  

(v) In the very low damping regimes, $l_{th} >> R$, the active particles randomly bounces off back and forth inside the cavity and occasionally escape the confinement hitting at the exit window. This exit mechanism works for the entire range of self-propulsion length. The escape rate in this regime is directly proportional to the pore number and their width; however, insensitive to the spacing between pores.     

Present study is motivated by the widespread interest in controlling the transport and diffusion of micro-swimmers in confined structures. Thinking of potential application in various systems with large variability of length and time scales, we explore all the regimes of damping, self-propulsion lengths (or rotational relaxation times) and translational diffusion. Therefore, we believe that the findings of this study can be used to design the most efficient artificial self-propelled Brownian tracer for targeted drug delivery and applications in natural and artificial devices.    

\appendix

\section{Derivation of  Equations (\ref{exit-eq-1}-\ref{exit-eq-2},\ref{2pore-1})}\label{A}

 When $l_{\theta} >> R$, as well as,  $l_{th} << R$, the active particle  preferably diffuses on the cavity wall.  Moreover, during diffusion along cavity circumference  self-propulsion velocity $\vec{v_0}$ remains parallel to the position vector $\vec{r}$. Simulation results presented in Fig.1(d,e) accord with this assertion.  Now exit problem from a single opening cavity is basically reduced to 1D diffusion between two absorbing points at $\delta \phi/2$ and $2\pi - \delta \phi/2$.  Thus, when the particle is positioned at the antipodal point and  $\vec{v}_0$ is radially directed at $t = 0$, the exit time is given,  
\begin{eqnarray}
\tau_{ex} = \frac{(2\pi - \delta\phi)^2 }{8D_\theta}=\frac{(2\pi - \delta\phi)^2 l_\theta}{8v_0}.\label{A1}
\end{eqnarray}
One the other hand, when the particles uniformly distributed over the cavity circumference with radially directed self-propulsion velocity at the starting point. Following Ref.~\cite{Gardiner}, the mean exit time is given by,  
\begin{eqnarray}
 \tau_{ex} =\frac{(2\pi - \delta\phi)^2 l_\theta}{12v_0} 
\label{A2}
\end{eqnarray}

Now consider the situation when particles are initially placed at the cavity centre  with uniformly distributed orientation of $\vec{v_0}$ over the range 0 to $2\pi$. Now following argument of exit mechanisms [see discussion related to Eq.(\ref{exit-eq-2}) in Sec III ], $(2\pi-\delta \phi)/2\pi$ fraction of particles exit the cavity with an average time,  
\begin{eqnarray}
\tau_{1}=  \frac{(2\pi - \delta\phi)^2 l_\theta}{12 v_0}+\frac{R}{v_0},\label{A3}
\end{eqnarray}
The rest of the particles, comprise $\delta \phi /2\pi$ fraction, exit with an average time $\tau_{2} = R/v_0$. Appropriate weight averaging of contribution from these two types of trajectories produces Eq.~(\ref{exit-eq-2}).
\begin{eqnarray}
\tau_{ex}=  \frac{(2\pi - \delta\phi)}{2\pi} \tau_{1} + \frac{ \delta\phi}{2\pi} \tau_{2} = \frac{(2\pi - \delta\phi)^3 l_\theta}{24\pi v_0}+\frac{R}{ v_0} \nonumber   \\ \label{A4} 
\end{eqnarray}

Lets us consider exit from a two opening  cavity when pore-centres are separated by an angle $\phi$ as shown in the Fig.~1(b). Following argument of exit mechanism in  Eq.~(\ref{2pore-1}): $(\phi - \delta \phi)/2\pi$ fraction of particles exit the cavity with an average time,   
\begin{eqnarray}
\tau_{1}=  \frac{(\phi - \delta\phi)^2 l_\theta}{12 v_0}+\frac{R}{ v_0}, \label{A5}     
\end{eqnarray} 
another $(2\pi-\phi - \delta \phi)/2\pi$ fraction
\begin{eqnarray}
\tau_{2}=  \frac{(2\pi-\phi - \delta\phi)^2 l_\theta}{12 v_0}+\frac{R}{ v_0}, \label{A6}     
\end{eqnarray}
Expressions for $\tau_1$ and $\tau_2$ are obtained based on the reasoning used in Eq.~(\ref{A2}).  The remaining $\delta \phi /\pi$ fraction exit with the average time, $\tau_{3}=R/v_0$. Appropriate weight average of exit times for the three kinds of trajectories yields Eq.(\ref{2pore-1}).

\section*{Acknowledgements}
 P.K.G. is supported by SERB Start-up Research Grant (Young
Scientist) No. YSS/2014/000853 and the UGC-BSR Start-Up Grant No.
F.30-92/2015. D.M. thanks SERB (Project No. ECR/2018/002830/CS), Department of Science and Technology, Government of India, for financial support and IIT Tirupati for the new faculty seed grant.
\section*{Data Availability}
The data that support the findings of this study are available within the article.
\section*{Conflict of interest}
The authors have no conflicts to disclose.

\end{document}